# High angular resolution coronography for adaptive optics[1]

Fabien Malbet

Laboratoire d'Astrophysique, Observatoire de Grenoble

BP 53, 38041 Grenoble cedex 9, France. E-mail: malbet@gag.observ-gr.fr

and

Jet Propulsion Laboratory, California Institute of Technology

MS 306-388, 4800 Oak Grove Drive, Pasadena, CA 91109, U.S.A.

## ABSTRACT

Recent adaptive optics systems in astronomy achieve high-angular resolution. With the extreme stability of the images, detection at very low fluxes can be reached using a coronograph at the diffraction limit of the telescopes. This paper is an overview of the issues of stellar coronography used at the diffraction limit. Image formation through such a system is illustrated by numerical simulations. The description of a coronograph implemented on the VLT adaptive optics prototype, COME-ON, is presented as well as the first observations.

*Subject headings:* instrumentation: miscellaneous – techniques: miscellaneous – stars: imaging – atmospheric effects

## 1. Introduction

Until the 1930's, the solar corona was only observed during total solar eclipses. To avoid this restriction, Lyot invented the solar coronograph (1930, 1931, 1939). This instrument was a solar telescope equipped with a mask, whose size was exactly that of the sun. The light coming from the sun disk was blocked, which allowed the corona, one million times fainter than the photosphere, to shine through. Later, the light rejection rate was improved by modifying slightly the original layout (Evans 1948; Newkirk & Bohlin 1963; Fort et al. 1978; Koutchmy & Belmahdi 1987).

Recently astronomers have tried to apply this method to observe stellar environments. By occulting the stellar luminous contribution, the circumstellar material becomes detectable. In 1984 a coronograph made of a reticle in the focal plane enabled astronomers to detect a circumstellar disk around the star $\beta$ Pictoris (Smith & Terrile 1984, Vilas & Smith, 1987). More recently, an infrared coronograph has been designed in order to detect brown dwarfs (Beuzit et al. 1991). The limiting factors for this kind of stellar coronograph are the lack of spatial resolution of the instruments and the very faint brightness of the stellar surroundings. To observe a planet separated

---

by less than 5 AU from a star located at 5 pc, spatial resolution better than 1″ is required. This is also the resolving power limit of present telescopes. To achieve a good efficiency, precise centering of the star on the coronographic mask is necessary. Due to atmospheric disturbance, this cannot be done efficiently unless the mask is large. Thus Clampin et al. (1992) have built an optimized coronograph suited to the good imaging of the ESO NTT telescope at La Silla (Chile).

A clear way to avoid atmospheric turbulence is to perform space-based observations. Following this idea, Bonneau et al. (1975) proposed a coronographic mode for the Faint Object Camera (FOC) of the Hubble Space Telescope. Breckinridge et al. (1982) also proposed a coronographic camera for the Space Telescope and Watson et al. (1991) suggested the use of a coronograph on a large aperture lunar-based telescope.

Another way to improve image quality is to use active optics techniques. The "Shift & Add" method used in real-time with a tip-tilt mirror, as developed by Durrance & Clampin (1989) at John Hopkins University then by Golimowski et al. (1992) and also for the camera HRCam at the Canada-France-Hawaii telescope (Racine & McClure 1989), improves the efficiency of a coronograph and enables reduction of the occulting mask diameter. Current adaptive optics systems can also be used. This is the approach chosen that will be described hereafter in this paper. With recent adaptive optics systems (Rigaut et al. 1991; Beckers 1993), the resolving power is limited by the telescope diffraction, less than 0.2″ in the near-infrared, and the resulting images are motionless. The optical star beam can be centered on the occulting mask with high precision and is available for standard coronographic processing. This design has been implemented in January and April 1991 on COME-ON (Malbet 1992a, 1992b), the VLT adaptive optics prototype at the ESO 3.6 meter telescope in La Silla (Chile). This system and the first observations are described in this paper. The paper also addresses the performance of the first version of this coronograph. In the summer of 1991, the prototype COME-ON was updated to COME-ON+: the optical scheme was changed and the coronograph removed. In parallel, Durrance & Clampin (1989) have designed a full adaptive optics system dedicated to coronographic imaging at John Hopkins University. Unfortunately the full system has not been built to date.

In order to estimate the science that can be achieved with high angular resolution coronography, the reader is referred to some representative papers published in classical stellar coronography: imaging of $\beta$ Pictoris disk (Smith & Terrile 1984; Paresce & Burrows 1987), circumstellar imaging of symbiotic stars, planetary nebulae (Paresce, Burrows & Horne 1988; Burgarella, Clampin & Paresce 1991; Burgarella & Paresce 1991, Robberto et al. 1993) and luminous blue variables (Paresce & Nota 1989; Burgarella & Paresce 1991), imaging of the supernova SN 1987A (Paresce & Burrows 1989). With the JHU Adaptive Optics Coronograph, stellar coronography has improved in spatial resolution, e.g. for imaging blue luminous variables (Nota et al. 1992; Clampin et al. 1993). Images with very high contrast have also been obtained around young stars (Nakajima & Golimowski 1995) and a very-low mass star has been detected around a binary (Golimowski et al. 1995). These results show the potential of a coronograph used on a full adaptive system. Recent results with the new coronograph of COME-ON+ by

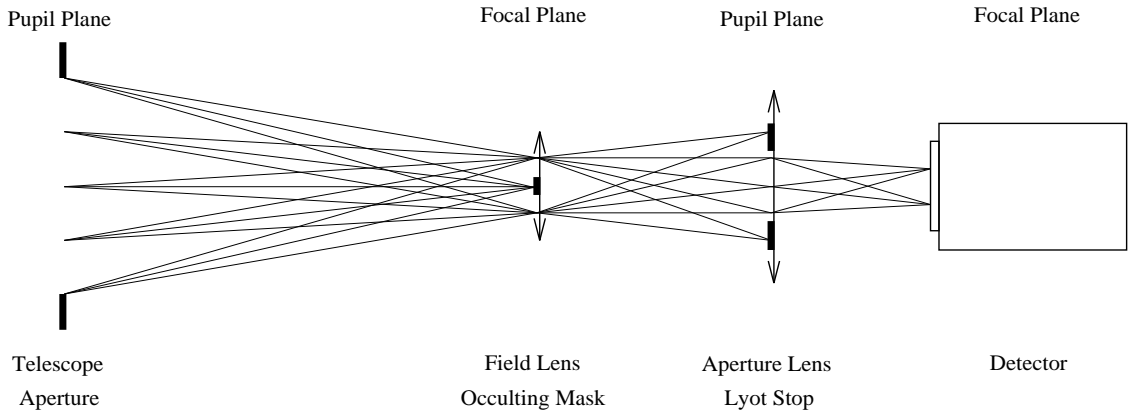

**Fig. 1.—** General optical schematic of a Lyot coronograph. Four optical planes are represented. Optical rays have been traced from 5 different points on the entrance pupil plane and for 3 different field rays.

Beuzit, Mouillet & Lagrange (1995) include imaging $\beta$ Pictoris disk up to $1''$ from the star in the near-infrared.

Image formation through a coronograph is treated in Sect. 2. Numerical simulations illustrate the performances in Sect. 3. Description of the coronographic mode of COME-ON and the first results constitute Sect. 4.

## 2. Image formation through a coronograph

The simplest coronograph is composed of an occulting mask introduced in the optical beam at the focus of the telescope. Lyot (1930, 1931) pointed out that the efficiency of a coronograph is appreciably increased, after the focal mask, by the use of a diaphragm in a pupil plane slightly smaller than the telescope aperture. A coronograph is then made of a focal occulting mask and a pupil stop, usually called Lyot stop. The mask is optically conjugated with the detector, and the Lyot stop with the telescope pupil (see Fig. 1).

In this section, I assume the adaptive optics system is perfect and restores a perfect wavefront. Therefore, I will consider the image formation in the far field approximation. Bonneau et al. (1975) and Labeyrie (1985) provided good presentations of this process, but without any mathematical formulation. Noll (1973) quantified the contribution of the Lyot stop for the intensity calculated on the optical axis. Wang & Vaughan (1988) have proposed an approximation to obtain the intensity in the focal plane in cases where a Lyot stop is used. The aim of this section is to describe the image formation through a coronographic system within the Fraunhofer diffraction theory, where the lenses are assumed perfect. In this section, I will show that the coronographic process is similar to a convolution, but with a point spread function different for each point of the sky.

A coronograph consists of cascaded optics. There is first an occulting mask which has two purposes:

- to reduce the light coming from the central star by a direct occultation

- to filter out the light at low spatial frequency.

The results are twofold:

- the on-axis stellar light in the following pupil plane is almost totally cancelled, depending on the size of the occulting mask,

- some light passes through. As the mask removes the low spatial frequencies in the image plane, the remaining light is on the edge of the pupil which corresponds to the highest frequencies.

The Lyot stop blocks the light located at the pupil edge. For a point located far from the central star, the occulting mask has a weak effect and the Lyot stop decreases the size of the aperture only slightly. Since the Lyot stop is located at a pupil plane, the removal of stellar light takes place everywhere in the focal plane, not only in the central part as with the occulting mask. In a way, the occulting mask is useful to avoid saturation of the detector with central stellar light, but not useful for detecting circumstellar material, because the diffraction pattern remains at the same level of intensity outside of the occulting mask. The purpose of the Lyot stop is to decrease this level of diffraction.

## 2.1. Instantaneous amplitudes

The amplitude of light arriving onto the focal mask is the convolution of the initial light amplitude of the object, $\psi_0(\vec{\Theta})$, with the normalized impulse response of the telescope, $G_T(\vec{\Theta})$:

$$\psi^m(\vec{\Theta}) = G_T(\vec{\Theta}) * \psi_0(\vec{\Theta}), \tag{1}$$

with $\vec{\Theta} = (\theta_x, \theta_y)$ is the position vector in terms of angular size.

The field amplitude just after the occulting mask is $\psi^m(\vec{\Theta})$ multiplied by $\amalg_M(\vec{\Theta})$, the transmission function of the mask. Therefore the resulting amplitude on the detector is:

$$\psi(\vec{\Theta}) = G_L(\vec{\Theta}) * \left\{ \amalg_M(\vec{\Theta}) \left[ G_T(\vec{\Theta}) * \psi_0(\vec{\Theta}) \right] \right\}, \tag{2}$$

where $G_L(\vec{\Theta})$ is the impulse response associated with the Lyot stop. By permuting the integrals of the two convolution operations, one finally gets:

$$\psi(\vec{\Theta}) = \int d\vec{\Theta}' \psi_0(\vec{\Theta}') \mathcal{G}(\vec{\Theta}'; \vec{\Theta}), \tag{3}$$

where

$$\mathcal{G}(\vec{\Theta}'; \vec{\Theta}) = \int d\vec{\Theta}_1 \, G_L(\vec{\Theta} - \vec{\Theta}_1) \, \Pi_M \, (\vec{\Theta}_1) G_T(\vec{\Theta}_1 - \vec{\Theta}'). \qquad (4)$$

The amplitude in the following pupil plane just before the Lyot stop, is the Fourier transform of the amplitude found just after the occulting mask:

$$\psi^L(\vec{f}) = \hat{\Pi}_M(\vec{f}) * [\Pi_T(\vec{f})\hat{\psi}_0(\vec{f})], \qquad (5)$$

where $\Pi_T$ is the pupil function (the Fourier transform of the impulse response $G_T$), $\hat{\Pi}_M$ is the Fourier transform of the mask and $\vec{f} = (u, v)$ is the spatial frequency coordinate. It can be written as:

$$\psi^L(\vec{f}) = \int d\vec{\Theta}' \psi_0(\vec{\Theta}')\mathcal{K}(\vec{\Theta}'; \vec{f}), \qquad (6)$$

where

$$\mathcal{K}(\vec{\Theta}; \vec{f}) = \int d\vec{\Theta}' \, \Pi_M \, (\vec{\Theta}') G_T(\vec{\Theta}' - \vec{\Theta}) e^{-i2\pi\vec{\Theta}'\cdot\vec{f}}. \qquad (7)$$

## 2.2. Intensity distribution

The intensity is defined as $I = \langle\psi\psi^\star\rangle$. In the case where the incident light is coherent,

$$\langle\psi_0(\vec{\Theta}_1)\psi_0^\star(\vec{\Theta}_2)\rangle = \psi_0(\vec{\Theta}_1)\psi_0^\star(\vec{\Theta}_2), \qquad (8)$$

one gets respectively in the focal plane and in the Lyot plane:

$$I(\vec{\Theta}) = \psi(\vec{\Theta})\psi^\star(\vec{\Theta}), \qquad (9)$$
$$\text{with } \psi(\vec{\Theta}) = \int d\vec{\Theta}' \psi_0(\vec{\Theta}')\mathcal{G}(\vec{\Theta}'; \vec{\Theta}), \qquad (10)$$
$$I^L(\vec{f}) = \psi^L(\vec{f})\psi^{L\star}(\vec{f}), \qquad (11)$$
$$\text{with } \psi^L(\vec{f}) = \int d\vec{\Theta}' \psi_0(\vec{\Theta}')\mathcal{K}(\vec{\Theta}'; \vec{f}). \qquad (12)$$

In the case where the incident light is incoherent,

$$\langle\psi_0(\vec{\Theta}_1)\psi_0^\star(\vec{\Theta}_2)\rangle = I_0(\vec{\Theta}_1)\delta(\vec{\Theta}_1 - \vec{\Theta}_2), \qquad (13)$$

where $I_0(\vec{\Theta})$ is the intensity distribution of the object and $delta(\vec{\Theta})$ is the delta-function. One gets respectively in the focal plane and in the Lyot plane:

$$I(\vec{\Theta}) = \int d\vec{\Theta}' I_0(\vec{\Theta}')\mathcal{F}(\vec{\Theta}'; \vec{\Theta}) \qquad (14)$$
$$\text{with } \mathcal{F}(\vec{\Theta}'; \vec{\Theta}) = \left|\mathcal{G}(\vec{\Theta}'; \vec{\Theta})\right|^2, \qquad (15)$$
$$I^L(\vec{f}) = \int d\vec{\Theta}' I_0(\vec{\Theta}')\mathcal{H}(\vec{\Theta}'; \vec{f}) \qquad (16)$$
$$\text{with } \mathcal{H}(\vec{\Theta}'; \vec{f}) = \left|\mathcal{K}(\vec{\Theta}'; \vec{f})\right|^2. \qquad (17)$$

## 2.3. Point Spread Function of a coronograph

The notion of point-spread-function (PSF) used in traditional optics is not valid in the context of a coronograph. The resulting intensity is not the convolution of the incident intensity by a PSF, because the modulation transfer function is no longer invariant to translation. An unresolved star will have a different response when located under the occulting mask than that it has when located outside the mask. However that is the intent of a coronographic system, to enhance low flux level while attenuating the unwanted axial light.

If the object is an unresolved star at position $\vec{\Theta}'$, the intensity distribution is $I_0(\vec{\Theta}) = I_0\delta(\vec{\Theta} - \vec{\Theta}')$, then

$$I(\vec{\Theta}) = I_0\mathcal{F}(\vec{\Theta}';\vec{\Theta}) \tag{18}$$

Therefore $\mathcal{F}(\vec{\Theta}';\vec{\Theta})$ is the "point spread function" associated with the direction $\vec{\Theta}'$. There is not a unique PSF but a PSF for every point on the object. This set of PSFs, $\mathcal{F}(\vec{\Theta}';\vec{\Theta})$, is easily calculated if one notices that it is the image of a point source at direction $\vec{\Theta}'$ through the system. Therefore $\mathcal{F}(\vec{\Theta}';\vec{\Theta})$ is only the PSF of the telescope occulted by the focal mask and convolved with the PSF of the Lyot stop.

In terms of discrete sampling, if the position on the detector is identified as $(i, j)$, one gets:

$$I_{i,j} = \sum_{k,l} I_{0k,l} F_{k,l,i,j}, \tag{19}$$

where $F_{k,l,i,j}$ is the 4D-tensor associated with the system transfer function. To recover the original $I_0$, the tensor $F$ has to be known and inverted.

The shape of $\mathcal{H}(\vec{\Theta}';\vec{\Theta})$ can be understood by the effect of diffraction of a star centered on the focal mask. This effect is to convolve the light distribution in the pupil plane by the mask impulse response (Bonneau et al. 1975; Labeyrie 1985). Since only the high spatial frequencies are not filtered by the occulting mask, the light is mainly suppressed far from the pupil edge. This results in a pupil which is over-intensified near the edges. For a star not centered on the optical axis, the focal mask has almost no effect. Therefore the light distribution in the pupil plane is not changed. The Lyot stop suppresses the bright edge fringe from the central bright star, but has minor consequences for an object far from the mask. The Lyot stop must be sized to the mask diameter so that the blocked area has a width equal to that of the mask impulse response in the pupil Lyot plane.

The coronograph PSF beyond the focal mask edge is very similar to the Lyot stop PSF. The Lyot stop PSF is obtained if the entrance pupil were the Lyot stop. Likewise the coronograph PSF far from the mask is obtained by removing the focal mask and not the Lyot Stop. In a complete coronograph, the detector intensity distribution is the convolution of (1) the intensity distribution just before the mask multiplied by the mask , and (2) the Lyot Stop PSF. When the off-axis direction is far from the mask edge by more than the width of the Lyot stop PSF, the intensity distribution is very close to the one obtained with no focal mask, but with a pupil.

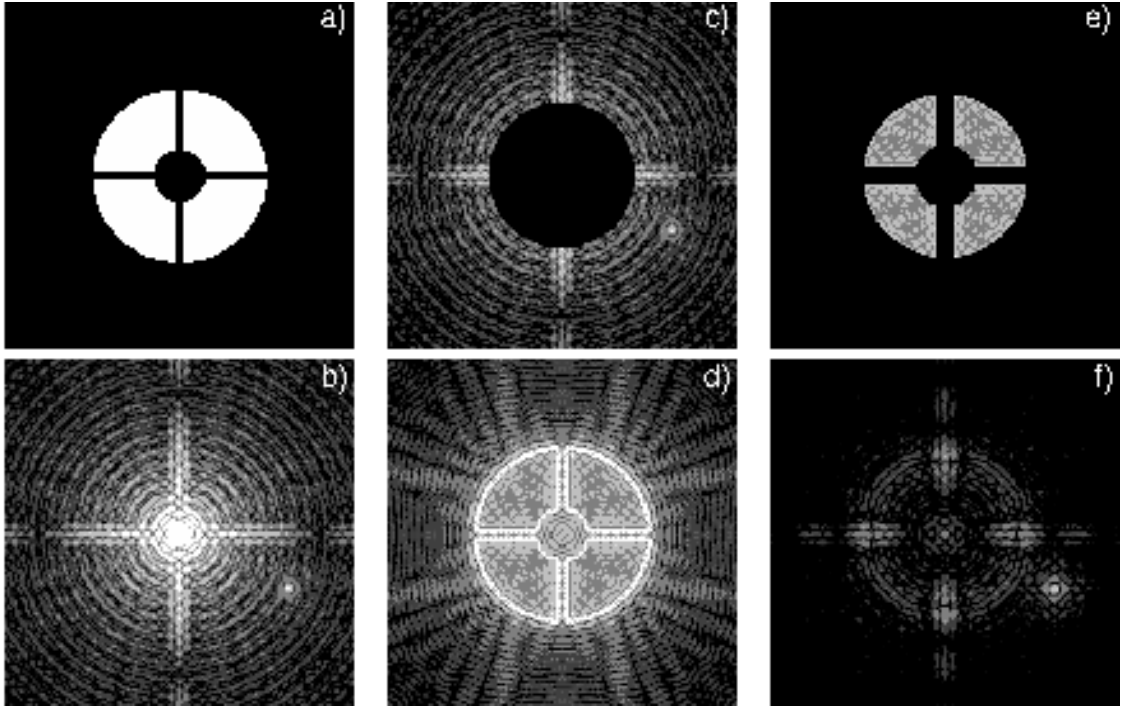

Fig. 2.— Intensity distribution in a coronographic system for detection of a binary star with a flux ratio of 1000 and a separation corresponding to the radius of about 20 bright Airy rings from the center. The focal mask stops the light inside 12 Airy rings. The panels show the intensity distributions: (a) in the entrance pupil plane, (b) in the first focal plane before the occulting mask, (c) in the first focal plane after the coronographic mask, (d) before the Lyot stop, (e) after the Lyot stop and (f) in the final focal plane of the detector. The logarithmic color scale is identical for the different images of the focal planes and also for the images of the pupil planes.

Figure 2 shows the different intensity distributions along the optical path of a coronograph in the case of a close binary. Note the over-intensity at the pupil edge in Panel (d) which has been removed in Panel (e) by using of a Lyot stop.

## 2.4. Using the Lyot stop

The Lyot stop decreases the intensity of the stellar diffraction wings at every point in the focal plane even far from the occulting mask. It modifies only slightly the intensity of an object located outside the coronographic mask. Therefore the contrast between the off-axis object and the level of intensity from the diffraction wings of the on-axis star increases. Without a Lyot stop the level of intensity of the stellar diffraction wings outside the occulting mask is not changed by the coronograph and the contrast between the object and the background level remains the same as it would be with no coronographic device. The contrast enhancement comes from the attenuation of the edge discontinuity in the pupil plane by the Lyot stop. This discontinuity size is inversely proportional to the occulting mask size (Fourier transform properties). Figure 3 shows the difference between a coronograph with a mask size large compared to the telescope PSF size (panel a) vs. a coronograph whose mask size is similar to the PSF size (panel b).

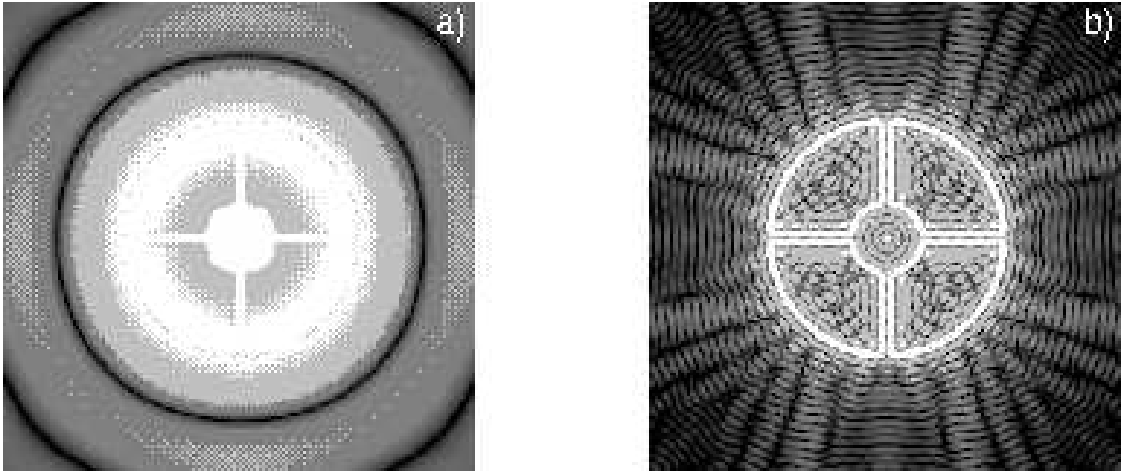

Fig. 3.— Intensity in the Lyot pupil plane for two kinds of coronograph: (a) occulting mask with a same size as the resolving power (1 Airy disk); (b) occulting mask size large compared to the resolving power (12 Airy rings). The figures display the intensity distribution in the pupil plane following the occulting mask. Note that the over-intensity due to the diffraction of the light on the edge of the occulting mask is sharper in panel (b) than in panel (a).

- If the mask size is large compared to the full width at half maximum (FWHM) of the telescope impulse response (Panel (a) of Fig. 3), then the over-intensity in the Lyot pupil plane will be sharp and may be suppressed by a Lyot stop slightly smaller than the pupil. This is the case for solar coronographs and also for stellar coronographs that use large masks compared to the FHWM of the image.

- If the mask size is of the same order of magnitude as the FWHM of the telescope PSF (Panel (b) of Fig. 3), then the over-intensity width is of the same order of magnitude as the pupil diameter. It is not possible to suppress it without significantly decreasing the Lyot stop size, resulting in a loss in resolution and collecting area. This is the case for coronographs used at high angular resolution (i.e. at the diffraction limit of the telescope), and the Lyot stop is not recommended.

If one is interested in the close neighborhood of an object, a small mask must be used. The Lyot stop will not significantly decrease the intensity of the central object. However if one is interested in objects further from the central object, the larger the focal mask, the sharper the over-intensity in the Lyot plane and the better the rejection rate.

## 2.5.  Summary

Image formation occurs similarly in a coronograph as in a simple optical instrument. However, the coronographic mask breaks up the system transfer function invariance. Therefore the point spread function depends on the object position in the field. The convolution operation is then a multiplication by a 4-fold tensor − the transfer function of the coronographic system. However, at a distance corresponding to the full width at half maximum (FWHM) of the Lyot stop impulse function from the focal mask, the coronograph PSF is similar to the Lyot stop PSF.

If the size of the coronographic mask is large compared to the telescope PSF, the coronograph has outstanding performance. In that case, over-intensity at the Lyot pupil edge is sharp and is easily stopped by a Lyot stop few percent smaller than the telescope entrance pupil. Obviously a coronograph works better with large masks. On the other hand, stellar coronography and studies of the circumstellar environment demand observations close to the star. Therefore very high angular resolution combined with coronography is required, but one must use a smaller mask (whose size is comparable to the FWHM of the PSF) and the Lyot stop is not as efficient.

The following section presents numerical simulations. Coronographic performances will be discussed in terms of the mask and Lyot stop size versus cutoff frequency of the telescope.

## 3.  Expected performances

In this section, I apply previous results to a simple case, and compute numerically the intensity at the Lyot plane, $I^L(\vec{f})$, and at the focal plane $I(\vec{\Theta})$. This section demonstrates the dependence of these quantities on both mask and Lyot stop size.

Consider the case of a circular pupil with cutoff frequency $f_c$, a circular occulting mask with angular diameter $\theta_m$, and a circular Lyot stop with cutoff frequency $\alpha f_c$ (with $\alpha > 0$). Furthermore, suppose the magnification between successive planes is one. The transfer functions and impulse responses used in the previous section are found in Table 1. As described at the end of Sect. 2.4., $\alpha$ should be such that the over-intensity in the Lyot pupil plane is suppressed. The diameter of the dark ring closest from the pupil edge has been chosen for the Lyot Stop diameter. Therefore there is a relation between $\alpha_1$ computed for occulting mask diameter $\theta_{m1}$ and $\alpha_2$ computed for occulting mask diameter $\theta_{m2}$:

$$\alpha_2 = 1 - (1 - \alpha_1)\frac{\theta_{m1}}{\theta_{m2}}. \qquad (20)$$

Table 1: Mathematical expressions for the transfer functions and impulse responses of the telescope entrance pupil, the mask plane, and the Lyot plane in the case of a circular pupil.

| Location | Direct Planes (focus) | Fourier Planes (pupil) |
|---|---|---|
| Telescope aperture | $G_T(\vec{\Theta}) = J_1(\pi f_c \theta)/\sqrt{\pi}\theta$ | $\Pi_T(\vec{f}) = \begin{cases} 2/\sqrt{\pi}f_c & \text{if } \rho < f_c/2 \\ 0 & \text{if } \rho > f_c/2 \end{cases}$ |
| Mask | $\amalg_M(\vec{\Theta}/\theta_m) = \begin{cases} 0 & \text{if } \theta < \theta_m \\ 1 & \text{if } \theta > \theta_m \end{cases}$ | $\hat{\amalg}_M(\vec{f}) = \delta(\vec{f}) - \theta_m\,J_1(\pi\theta_m\rho)/2\rho$ |
| Lyot stop function | $G_L(\vec{\Theta}) = \alpha f_c J_1(\pi\alpha f_c\theta)/2\theta$ | $\Pi_L(\vec{f}) = \begin{cases} 1 & \text{if } \rho < \alpha f_c/2 \\ 0 & \text{if } \rho > \alpha f_c/2 \end{cases}$ |

with $\theta = ||\vec{\Theta}|| = \sqrt{\theta_x^2 + \theta_y^2}$ and $\rho = \sqrt{u^2 + v^2}$

The optical analysis software, COMP (Controlled Optics Modeling Package, Redding et al. 1992), was used to calculate the intensity distribution at different planes in the coronograph.

One may wonder whether results computed with simple assumptions (circular pupil and circular Lyot stop) are useful for understanding actual systems, since telescopes usually have secondary obscuration and spider? The final performance is slightly different, but the effects are basically identical: the main effect comes from the diffraction of light at the edges of the pupil, either at the outer edge of the pupil, at the edge of the secondary obscuration or at the edge of the spider. A circular pupil makes computing but also interpretation of the results easier.

### 3.1. Impulse responses of a coronograph at the Lyot stop and focal planes

Figure 4 and Fig. 5 display the functions $\mathcal{H}(\vec{\Theta}';\vec{f})$ and $\mathcal{F}(\vec{\Theta}';\vec{\Theta})$. A squared mesh of $1024 \times 1024$ pixels was used. The FWHM is equal to 32 pixels $(1/fc)$. The PSFs are computed for two mask diameters, 39 pixels corresponding to the first dark ring and 104 pixels corresponding to the third dark ring of the telescope PSF. The radial variations of $\mathcal{F}(\vec{\Theta}';\vec{\Theta})$ and $\mathcal{H}(\vec{\Theta}';\vec{f})$ are plotted for three PSF positions corresponding to:

- the optical axis (Panels $a$ and $d$)

- the third bright ring of the Airy pattern centered on the optical axis (Panels $b$ and $e$)

- the fifth bright ring of the Airy pattern centered on the optical axis (Panels $c$ and $f$)

The effect of the coronograph is demonstrated in Panels $a$ and $d$ of Fig. 4 and Fig. 5. In Fig. 4, one can see the over-intensity surrounding a dark pupil. It is not the case for the off-axis directions

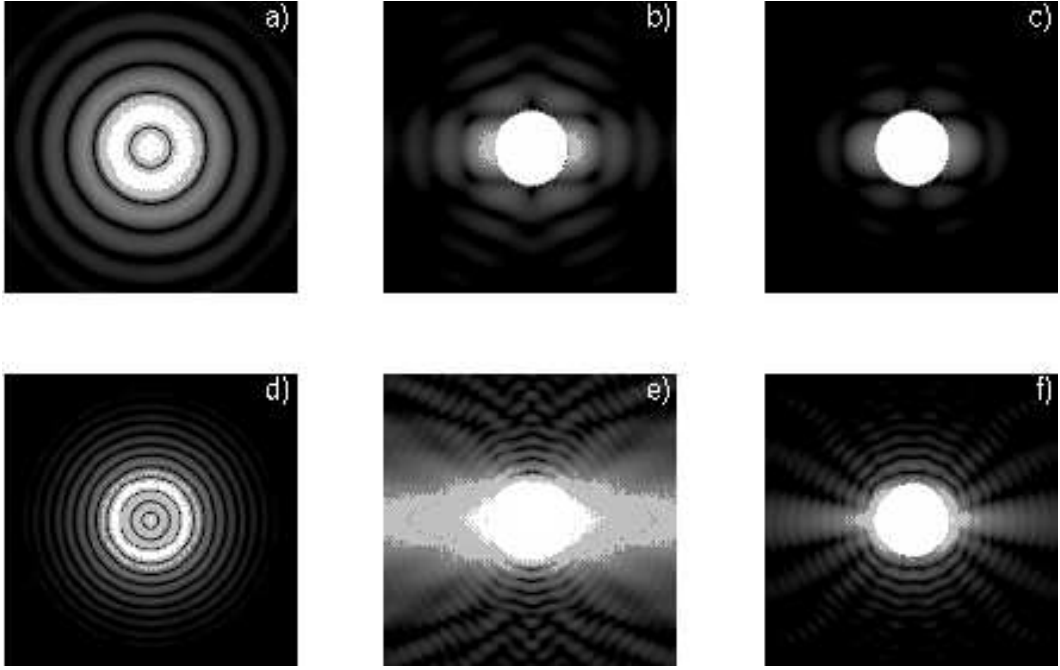

Fig. 4.— Pupil plane impulse response $\mathcal{H}(\vec{\Theta}';\vec{f})$ of a coronograph with a circular entrance pupil. The scale is identical in the 6 panels.

(a), (b), (c) corresponds to a coronograph whose focal mask has the same diameter as the first Airy dark ring. (c), (d), (e) belongs to a coronograph whose focal mask has the same diameter as the third dark Airy ring.

(a) and (d) are the intensity distributions in the Lyot plane for $\vec{\Theta}' = 0$. (b) and (e) are the intensity distributions in the Lyot plane for $\vec{\Theta}'$ corresponding to the third dark Airy ring. (c) and (e) are the intensity distributions in the Lyot plane for $\vec{\Theta}'$ corresponding to the fifth dark Airy ring.

(Panels $b, c, e, f$). The difference between Panel $a$ and $d$ is the width of the over-intensity due to the size of the occulting mask. In Fig. 5, the influence of the size of the occulting mask. First the spatial resolution is less in upper panels compared to lower panels. Secondly the on-axis direction does not present a volcano shape and the central peak is dominant. However with both occulting mask sizes, the gain is evident in intensity. The color scale is logarithmic and the intensity for the off-axis direction is much more bright than for the on-axis direction. One can also notice that the effect of the mask is almost canceled for the directions far from the mask edge (Panels $b, c, f$). Quantitative results are given in the following section.

## 3.2.   Rejection rate and intensity gain

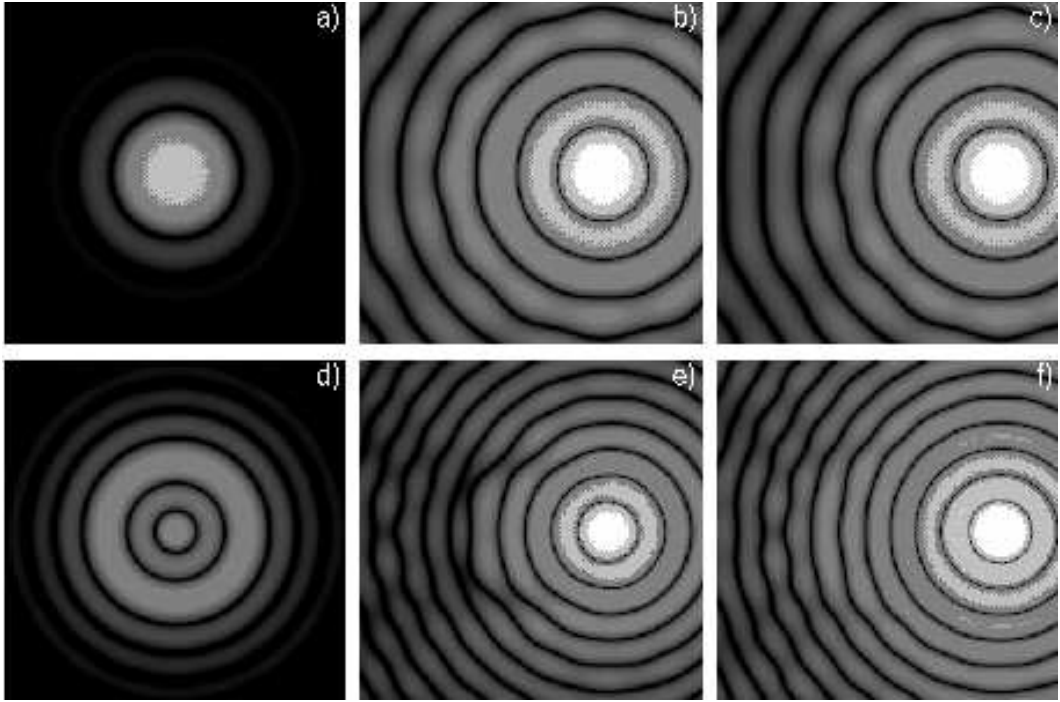

Fig. 5.— Detector plane impulse response $\mathcal{F}(\vec{\Theta}'; \vec{f})$ of a coronograph with a Lyot stop ($\alpha \neq \infty$). The scale is identical in the 6 panels.

(a), (b), (c) corresponds to a coronograph whose focal mask has the same diameter as the first Airy dark ring. (c), (d), (e) belongs to a coronograph whose focal mask has the same diameter as the third dark Airy ring.

(a) and (d) are the intensity distributions in the focal plane for $\vec{\Theta}' = 0$. (b) and (e) are the intensity distributions in the focal plane for $\vec{\Theta}'$ corresponding to the third dark Airy ring. (c) and (e) are the intensity distributions in the focal plane for $\vec{\Theta}'$ corresponding to the fifth dark Airy ring.

The rejection rate $R$ of a coronograph is by definition

$$R = \frac{I_{w/o}}{I_w},\tag{21}$$

where $I_w$ is the total intensity of the star in presence of coronograph and $I_{w/o}$ is the total intensity of the star in absence of coronograph. It depends on the mask and Lyot stop sizes as well as the angular separation between the central star and the interesting object. Figure 6 displays the rejection rate as a function of the distance from the optical axis for 3 different mask diameters (first, third and fifth dark rings). The mesh is the same as the one used in previous section.

Table 2 gives the result of the asymptotic rejection rate for a coronograph equipped with Lyot stop and a coronograph without one. Improvement is clear between the two designs. However the gain is less when the mask is small because the Lyot stop suppresses a large part of the pupil. The third column gives an estimate of the gain in magnitude on the stellar profile due to the addition of the Lyot stop. $\Delta m$ is equal to $2.5 \log(2R_w/R_{w/o})$ and is the same everywhere in the focal plane, except near the occulting mask. The factor 2 comes from the new repartition of the light after the convolution due to the Lyot mask. About 50% of the light is diffracted in the rings inside the mask area. The numbers can be compared with the actual results obtained on the sky and presented in the next section.

Table 2: Expected rejection rates of a stellar coronograph at high spatial resolution

| Mask diameter | with Lyot stop | w/o Lyot stop | $\Delta m$ |
|---|---|---|---|
| $1^{\text{st}}$ ring | 20 | 6 | 2.0 |
| $2^{\text{nd}}$ ring | 66 | 11 | 2.6 |
| $3^{\text{rd}}$ ring | 120 | 16 | 2.9 |
| $4^{\text{th}}$ ring | 180 | 21 | 3.1 |
| $5^{\text{th}}$ ring | 250 | 26 | 3.2 |
| $10^{\text{th}}$ ring | 580 | 52 | 3.4 |

### 3.3. Expected limitations from imperfect actual systems

Adaptive optics systems are experimentally not perfect. They only correct a certain number of modes. For example, tip-tilt compensation systems deal with only two modes. The COME-ON system, described in the next section, corrects 21 modes and the VLT adaptive optics is planned to correct about 256 modes. Not all the modes are useful depending on the wavelength and the telescope diameter. At $\lambda = 10\mu m$, only tip-tilt correction is necessary for a 4-m telescope in a good site. With 21 actuators on a 3.6-m telescope, COME-ON provides almost full correction for wavelengths larger than the near-infrared $K$ band (Rigaut et al. 1991). $K$ band images themselves are not fully diffraction-limited. The limit between full correction and partial correction is given by Lord Rayleigh's criterion (Maréchal & Françon 1970; Mariotti 1988): residual phase errors must be smaller than $\lambda/4$.

In the case of full correction, aberrations are small phase errors ($\Delta\phi \ll \pi/2$) and can be understood as scatter-type aberrations. The phase contribution $e^{i\Delta\phi}$ becomes $1 + i\Delta\phi$ and the scattered light is linearly added to the image. The intensity of the scattered light is proportional to the source intensity in the image. The pattern corresponds to the Fourier transform of the phase errors convolved with the PSF. This phenomenon is similar to the scattered light which comes from the collecting mirrors of large telescopes and which limits the dynamical range of spatial telescopes (Brown & Burrow 1990; Malbet, Yu & Shao 1995). The coronographic image is then photon-noise limited by the scattered light and many exposures are needed to decrease the background noise by subtraction. This limitation does not originate in the coronograph, but in the adaptive optics system correction.

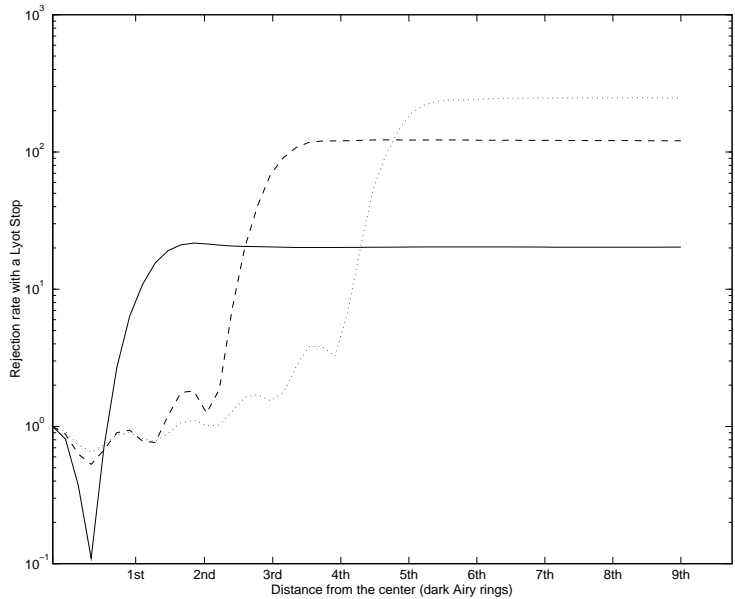

Fig. 6.— Coronograph rejection rate with Lyot stop. The diameter of the different masks are respectively: first dark ring (solid line), third dark ring (dashed line) and fifth dark ring (dotted line). The $x$-axis represents directions at different angular distance from the optical axis. The size of the Lyot stop has been computed to match the first dark ring at the pupil edge.

At shorter wavelengths, adaptive optics systems work in partial correction and are no longer diffraction-limited. Image formation becomes much more complex. We try to provide some answers but the understanding of this phenomenon is still limited. Beckers & Goad (1987) have shown that long-exposure stellar profiles in partial correction are spiked helmet-shaped in which an Airy shaped spike is superposed on a broad halo-like background with a width approximately equal to that of the original seeing disk. This was confirmed experimentally by Rigaut et al. (1992). As a matter of fact, what happens in the image formation through a coronograph is difficult to understand without extensive simulations and experimental tests. Qualitatively, one can estimate that the Airy shaped spike behaves like the fully corrected image and the broad halo like the seeing-limited image. The main limitation would come from the halo as in classical coronography.

Table 3: Rejection rate of coronographs with no accurate mask centering.

| Mask diameter | with Lyot stop | | without Lyot stop | |
| (centering error) | $(\theta_m/10)$ | $(\theta_m/2)$ | $(\theta_m/10)$ | $(\theta_m/2)$ |
| --- | --- | --- | --- | --- |
| 1st ring | 17 | 15 | 6 | 3 |
| 2nd ring | 66 | 42 | 11 | 7 |
| 3rd ring | 120 | 47 | 16 | 9 |
| 5th ring | 250 | 49 | 21 | 12 |
| 10th ring | 580 | 54 | 52 | 17 |

An interesting case is when the star is not accurately centered on the coronographic mask. This is relatively usual since in actual systems the star is never very accurately centered on the mask. Let us assume a $\sigma$-wide gaussian distribution of the centering probability and that the fluctuations of the rejection rate comes only from the centering of the central bright star. Table 3 presents the numerical results of this model by using the same procedure as in Sect. 3.2. to calculate the rejection rate. Rejection rates from a bad centering of one tenth of the mask size are not too degraded. For a bad centering of one half mask size, the rejection rates decrease by a factor of 2 (without a Lyot stop) up to a factor of 10 (with a Lyot stop). This is the reason why tip-tilt compensation system gives better results than classical stellar coronography. However the performance of tip-tilt compensated systems is still limited by the width of the stellar profile.

Another potential limitation is the isoplanatic angle. The isoplanatic angle is a radius of a circle on the sky over which atmospheric wavefront disturbances are practically identical. If the averaged altitude of the seeing layer is 6 km, the isoplanatic angle is about $60''$ at $\lambda = 2.2\mu m$ and about $3''$ at $\lambda = 0.5\mu m$. High angular resolution coronography with diameter of telescope larger than 2 m have always a spatial resolution much smaller than the isoplanatic angle and is likely to be limited by this effect.

## 4. The coronographic mode of COME-ON

### 4.1. Description

COME-ON is a prototype astronomical adaptive optics system for the VLT. It has been designed and built in France (Observatoire de Paris, ONERA, and Laserdot) in collaboration with the European Southern Observatory (ESO). The instrument design and performance have been described by Rousset et al. (1991) and Rigaut et al. (1991).

The COME-ON optical design permits the insertion of a stellar coronograph. The stellar beam is split by a dichroic into a visible beam used for wavefront sensing and an infrared beam used for science. This infrared beam is corrected by the deformable mirror located before the dichroic beam-splitter. The outcoming beam is corrected for the atmospheric turbulence and the images are very stable. The coronograph design is inserted between the last pupil plane and the focus. This design allows an easy implementation for astronomical use.

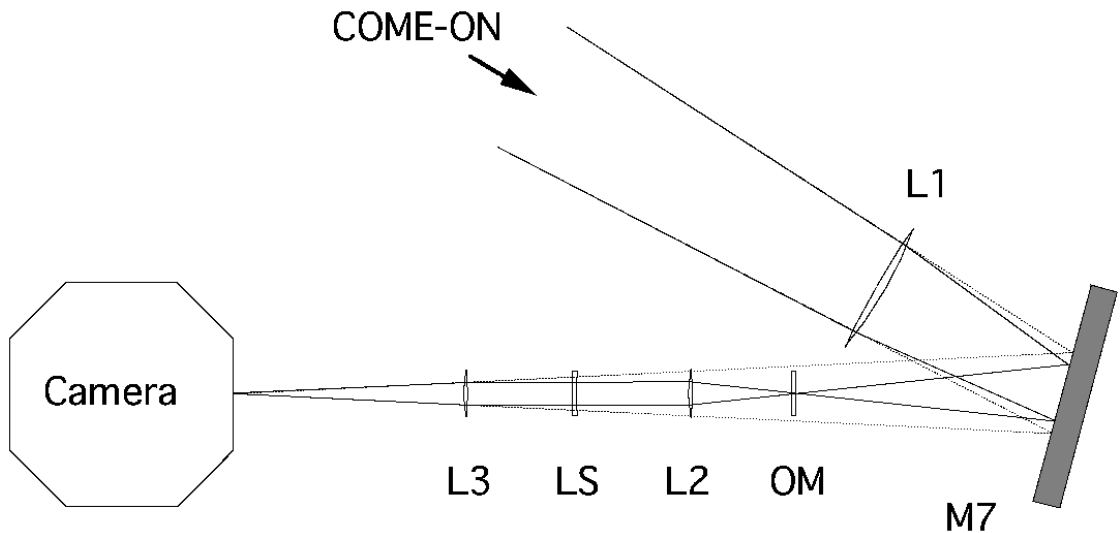

Fig. 7.— Optical layout of COME-ON coronographic mode. The solid lines represent the envelope of the optical beam when the coronograph is inserted. The dashed lines represent the original optical path of the light coming from COME-ON when the coronograph is not used.

### 4.1.1. Optical layout

The optical layout is shown in Fig. 7, where only the imaging arm of COME-ON is drawn. In order to implement all coronographic features, focal and pupil planes are created for the focal mask OM and the pupil Lyot stop LS. Three lenses L1, L2 and L3 are used to create these planes:

- L1 reduces the focal distance and forms the star image on the occulting mask OM. The optical aperture is then f/20 instead of the original f/40.

- OM is a glass plate that supports the occulting masks,

- L2 creates a plane conjugate to the entrance pupil plane,

- LS is a glass plate that supports the Lyot stop,

- L3 forms a convergent beam identical to the original beam and images the masked star onto the detector. The optical aperture is then f/40 or f/80 depending on the camera optical configuration.

These 5 optical elements are installed on an optical plate inserted between the On-Off mirror of COME-ON (M6 in the quoted publications) and the infrared detector. The M7 plane mirror only changes the beam direction to optimize space utilization. The lenses and glass plates which support the mask and Lyot stop are made from insoluble Fluoride glass (AFG 450 made by Le Verre Fluoré) which transmits infrared wavelengths.

### 4.1.2. Occulting masks and Lyot stop

The masks were made using microphotolithographic techniques at LARCA, a laboratory of the Observatoire de Paris. The 5 step procedure is:

- evaporation of chromium and then gold on the glass,

- deposition of a photosensitive resin layer

- ultraviolet illumination of the resin through a mask with the desired pattern,

- chemical removal of the chromium and the gold in the open zones,

- ultraviolet illumination of the rest of the glass support to remove the remaining resin.

To allow for different types of observation, 8 masks of differing diameters were placed on the glass support. A remote controlled translation stage shifted from one mask to the next and positioned the mask over the star. The mask sizes were chosen to approximately fit the diameter of the first dark diffraction rings at the different infrared wavelengths J, H, K, L and M. The chosen sizes were:

$$95, 120, 175, 205, 215, 250, 370, 460 \ \mu m$$

,

corresponding on the sky to $0\rlap{.}''27$–$1\rlap{.}''3$.

The Lyot Stop was made with the same technique. Only one stop was used. It was designed proportionally to the pupil shape: a circular aperture with a secondary obscuration and 4 spider arms. The pupil was reduced by 10% in size in order to increase the rejection rate. The glass support was mounted on a translation stage allowing fine positioning of the stop.

Table 4: Elementary integration for the observed star and performance of the coronograph

|  | Elementary integration time | | Gain | Gain | Rejection |
|---|---|---|---|---|---|
|  | with coron. | w/o coron. | in time | in intensity | rate |
| Sirius | 192 ms | 16 ms | 12 | 9.73 | 51 |

## 4.2. Observations

Observations using the coronographic mode on COME-ON were conducted January 1991 and April 1991 with the ESO 3.6-m telescope during nights dedicated to young stellar object observations. Only preliminary tests were performed to qualify the system for future use.

Images were recorded with a $32 \times 32$ InSb IR camera array (Lacombe et al. 1989). The pixel scale was $0''.056$ on the sky using the f/80 camera configuration. Chopping allowed for alternating on-source and sky exposures. Standard dead pixel elimination, flat fielding, and specific cross-talk correction between lines were performed for each object and sky elementary frame. The sky was subtracted from each object elementary frame, and the frame's total intensity was normalized.

One star was observed: Sirius. The images were taken in the infrared K band. This star is bright and the integration time was set at the detector saturation limit. Twenty elementary images were added together to create an image. The optical layout for the coronographic image used an occulting mask and a Lyot Stop. For the image without the coronograph, the occulting mask was taken out, but the Lyot stop remained. Elementary integration times with and without the coronograph are summarized in Table 4. The spatial achieved resolution has been estimated at $0.3''$. This resolution is not diffraction-limited on this system because at this wavelength the wavefront correction is only partial, but the resolution is much better than any previous coronographic observation. Quantitatively, Rigaut et al. (1991) gives a residual phase error of $11$ rd$^2$ for $\lambda = 0.5\mu m$, which corresponds to an absolute phase error of $0.3\mu m$ or $\lambda/3.7$ in the $K$ band., slightly smaller than the Rayleigh's criterion.

The size of the mask used was $175$ $\mu m$ corresponding to $0.5''$. It was a little smaller than the second dark Airy ring, and with the achieved resolution, occulted the central core of the PSF. The smallest coronographic mask used previously in other coronographs is $4''$ (Golimowski et al. 1993).

## 4.3. Results

In this section, I describe the specifics of coronographic reduction process and I provide the technical results of the observations. In addition of the calculation of the rejection rate, I show that the coronograph does not destroy the star radial profiles beyond a radius equal to the sum of the mask radius and the FWHM.

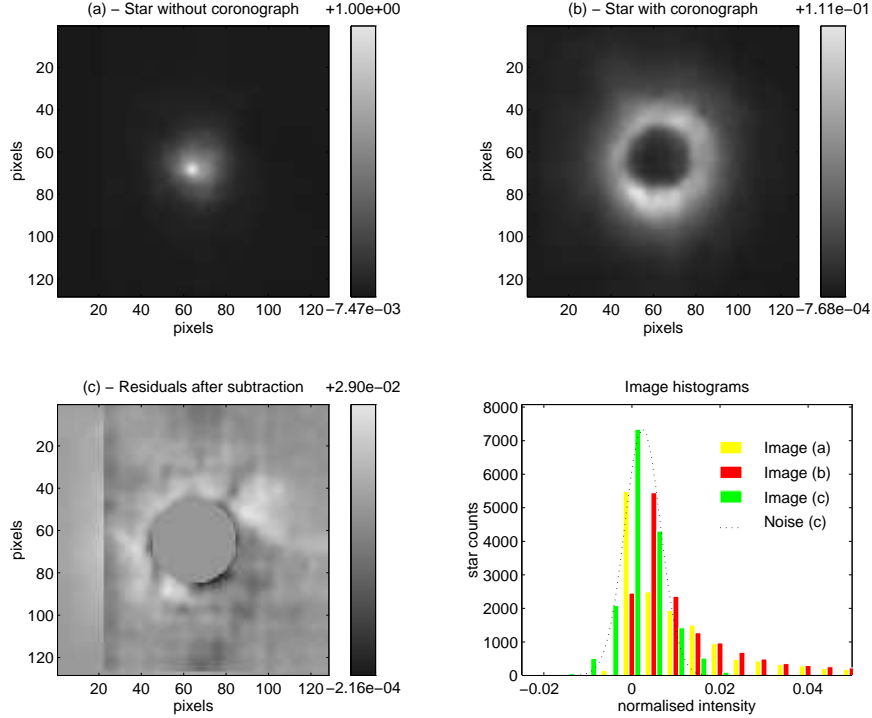

Fig. 8.— Interpolated images of Sirius (a) without and (b) with the COME-ON coronograph. The elementary integration time has been multiplied by 12 between the two images to reach the same level of flux. The intensities have been normalized in function of the intensities in image(a). Image (c) is the image of the residuals from the subtraction of image (a) from image (b). The histogram shows that the residuals in image (c) are smaller than the background noise level in images (a) and (b). Sirius was close to the edge of the detector in image (a) producing the "white" vertical band in subtracted image (c). See text for detailed comments on data processing.

Upper panels of Fig. 8 show the images of Sirius taken with and without the coronograph. The dynamic range was about the same in both series of images but the elementary time was multiplied by a factor 12 for the coronographic data. This gives an estimate of 12 for the gain in intensity. A better estimate is given further.

The coronographic reduction process consists in subtraction of the PSF from the coronographic image. In this paper the PSF is the same star observed with the coronograph but without the occulting mask (however the Lyot stop remains). Therefore I show that after accurate centering, subtraction of the PSF image from the coronographic image gives only background noise. The different steps of the process are:

- Linear interpolation of images (a) and (b) by a factor of 4 in order to get enough precision in the centering.

- Computation of software mask by convolving a hard-edge circular mask of $0.5''$ centered on the centroid of image (a) with the PSF (image (a)).

- Multiplication of the convolved mask by the hard-edge mask in order to suppress the weight of the central pixels in the cross-correlation.

- Cross-correlation of image (a) multiplied by the software mask and image (b) in order to know the angular separation between images (a) and (b). This method has the advantage to use only pertinent information (the intensity distribution outside of the occulting mask) and is unaltered from the remaining light at the mask.

- Translation of image (a) by a distance corresponding to the calculated separation.

- Computation of the ratio $\alpha$ between the two images by a least-square method. The $\chi^2$ metric is given by:

$$\chi^2 = \sum_{i,j} (b_{ij} - \alpha a_{ij})^2 M_{ij} \tag{22}$$

where $a_{ij}$ and $b_{ij}$ are the elements of image (a) and image (b) padded with zeros and $M_{ij}$ are the elements of the software mask used for the cross-correlation and centered on the star.

- Image (c) is the result of image (b) minus $\alpha$ times image (a), then multiplied by the software mask.

The results obtained for Sirius are shown in Fig. 8. The coefficient $\alpha$ is 9.73 corresponding to the gain in intensity. The rejection rate for the star is calculated by dividing the flux in image (a) by the flux in image (b) and is 51. The rejection rate for Sirius is very close to the theoretical value calculated in Table 2 where the rejection rate for an obscuration two Airy rings wide is 66 for a coronograph with a Lyot stop. The difference can be attributed to misalignment errors and to optical bench flexures. To show the performances of the system, a histogram of the three images has been computed in Fig. 8 (Panel $d$). One can see that the residuals (image (c)) are below the background noise floor of image (a) and image (b). This means that the residuals from the subtraction are at the noise floor. This noise floor is reached mainly because of the noise level in image (a).

The reduction process for coronography at the diffraction limit described in this section can be adjusted, but one still needs calibration images of the PSF (used to smooth the hard-edge mask). If a better signal-to-noise ratio is desired, a coronographic image of the PSF is required in order to get a smaller noise level. Then the coronographic image of the PSF can be subtracted from the coronographic image of the object.

In order to convert this result in useful astronomical units one can say the result corresponds to a gain of $2.47 \pm 0.02$ in magnitude for a mask size of $0.5''$. At a distance of $1''$ from the center of the mask, the level of star diffraction is then $12.5$ magnitude fainter than the peak of flux from the star.

## 5. Conclusion

This paper presents an overview of the coronographic process for high angular resolution imaging. It shows high angular resolution coronography is a powerful technique that can reveal circumstellar environments.

The occulting mask and Lyot stop must be carefully chosen. A relationship exists between them and the instrument resolution. The larger the occulting mask is relative to the PSF, the smaller the pupil reduction of the Lyot stop. If the mask is close to the PSF size, a Lyot stop is not necessarily required for it does not improve the rejection rate and only reduces spatial resolution. Simulations show the coronographic impulse functions in different planes and illustrate the coronograph principle. In the case of bad centering, the image data are still valuable, but the dynamic range and the rejection rate are lower. Therefore accurate centering is important for efficient observations and a better signal-to-noise ratio, but does not degrade data. A coronograph mode has been implemented on the adaptive optics prototype COME-ON. Initial observations indicate the coronograph has the expected behavior and illustrate the potential of such a system for astronomical observations.

I would like to thank the COME-ON team, the Département de Recherche Spatiale (DESPA) of the Observatoire de Paris and the Laboratoire d'Astrophysique of the Observatoire de Grenoble for supporting this work. I am especially grateful to Kent Wallace and the Spatial Interferometry Group at JPL, Jean-Luc Beuzit, Jérôme Bouvier and the referee for giving comments. Data are based on observations obtained at the European Southern Observatory, La Silla, Chile

---